\documentclass[runningheads,a4paper]{llncs}
\usepackage{amsmath,amssymb}
\usepackage{bbm}    % Nicer symbols for sets
\usepackage{mma}    % Mathematica typesetting environment
\usepackage{url}    % For displaying e-mail addresses

\def\D#1{D_{\!#1}}
\def\S#1{S_{\!#1}}
\def\d{\partial}

\def\Q{\mathbbm Q}

\def\F{\mathbbm F}

\def\O{\mathbbm O}
\def\rmd{\,\mathrm{d}}  % for integrals
\def\Ann{\operatorname{Ann}}

\renewcommand{\leq}{\leqslant}

\def\pow{\mathbin{\raisebox{-2.5pt}{\hbox{\large$\hat{}$}}}}  % Power x^y

\renewcommand{\labelenumi}{(\arabic{enumi})}

\urldef{\mailck}\path|christoph.koutschan@ricam.oeaw.ac.at|
\urldef{\mailpp}\path|peter.paule@risc.jku.at|
\urldef{\mailss}\path|sergei@asu.edu|

\makeatletter
\renewcommand\@makefntext[1]{\leftskip=2em\hskip-2em\@makefnmark#1}
\makeatother

\begin{document}

\mainmatter

\title{Relativistic Coulomb Integrals and Zeilberger's Holonomic Systems Approach II}
\titlerunning{Relativistic Coulomb Integrals and the Holonomic Systems Approach II}
 
\author{Christoph~Koutschan\inst{1}\and Peter Paule\inst{2}\and Sergei K. Suslov\inst{3}}
\institute{
        Johann Radon Institute for Computational and Applied Mathematics (RICAM),\\
        Austrian Academy of Sciences\\
        Altenberger Stra\ss e 69, A-4040 Linz, Austria\\
        \mailck\and
        Research Institute for Symbolic Computation (RISC)\\
        Johannes Kepler University\\
        Altenberger Stra\ss e 69, A-4040 Linz, Austria\\
        \mailpp\and
        School of Mathematical and Statistics Sciences \\
        Arizona State University\\
        Tempe, AZ 85287-1804, U.S.A.\\
        \mailss}

\maketitle

\begin{abstract}
  We derive the recurrence relations for relativistic Coulomb integrals
  directly from the integral representations with the help of computer algebra
  methods. In order to manage the computational complexity of this problem, we
  employ holonomic closure properties in a sophisticated way.
  \footnote{The final publication is available at link.springer.com.\\
    DOI 10.1007/978-3-642-54479-8\_6}
\end{abstract}


\section{Introduction}
\label{sec.intro}

This work was originally motivated by experimental and theoretical progress in
checking Quantum Electrodynamics in strong fields
\cite{Gum05,Gum07,ShabGreen,ShabYFN08} (see also the references therein).  A
study of the expectation values of the Dirac matrix operators multiplied by
the powers of the radius between the bound-state relativistic Coulomb wave
functions was initiated in~\cite{SuslovJPB09,Suslov10} and continued, from
computer algebra point-of-view, in \cite{PauleSuslov13}.

We present the radial wave functions~$F$ and~$G$ in the following form:
\begin{equation}\label{eq.FGdef}
  \begin{pmatrix}F(r)\\[1ex] G(r)\end{pmatrix} =
  E(r) \begin{pmatrix}\alpha_1 & \alpha_2\\[1ex] \beta_1 & \beta_2\end{pmatrix}
  \begin{pmatrix}L^{(2\nu)}_{n-1}(2a\beta r)\\[1ex] L^{(2\nu)}_n(2a\beta r)\end{pmatrix}
\end{equation}
where the prefactor~$E$ is given by
\[
  E(r) = a^2 \beta^{3/2} \sqrt{\frac{n!}{\gamma\,\Gamma(n+2\nu)}}\, (2a\beta r)^{\nu-1} e^{-a\beta r}
\]
and where $L_n^{(\lambda)}(x)$ denotes the Laguerre polynomials. The quantities
$\alpha_1$, $\alpha_2$, $\beta_1$, and $\beta_2$ are given by the following
expressions:
\begin{align}
  \alpha_{1,2} & = \pm \sqrt{1+\varepsilon} \, \big((\kappa-\nu)\sqrt{1+\varepsilon} \pm \mu\sqrt{1-\varepsilon}\big), \label{eq.alphas}\\
  \beta_{1,2} & = \sqrt{1-\varepsilon} \, \big((\kappa-\nu)\sqrt{1+\varepsilon} \pm \mu\sqrt{1-\varepsilon}\big). \label{eq.betas}
\end{align}
The symbols $a$, $n$, $\beta$, $\varepsilon$, $\kappa$, $\mu$, and $\nu$
denote physical constants and they are connected by the following relations:
\begin{align}
  \kappa ^2 & = \mu ^2+\nu ^2, \label{eq.rel1}\\
  a^2 & = 1-\varepsilon ^2, \label{eq.rel2} \\
  \varepsilon\mu & = a(\nu +n). \label{eq.rel3}
\end{align}
We are interested in computing the relativistic Coulomb integrals of the
radial wave functions where $p$ is a non-negative integer:
\begin{align}
 A_p & = \int_0^\infty r^{p+2} \big( F(r)^2 + G(r)^2 \big) \rmd r, \label{eq.A} \\
 B_p & = \int_0^\infty r^{p+2} \big( F(r)^2 - G(r)^2 \big) \rmd r, \label{eq.B} \\
 C_p & = \int_0^\infty r^{p+2} \big( F(r) G(r) \big) \rmd r. \label{eq.C}
\end{align}

The title of the present paper bears the attribute ``II'' which refers the
reader to our first study of applying the holonomic systems approach to
relativistic Coulomb integrals: in~\cite{PauleSuslov13} the desired
recurrences for the integrals $A_p$, $B_p$, and $C_p$ were derived starting
from hypergeometric series representations \cite[(8)-(10)]{PauleSuslov13}
of these integrals.  In order to obtain such representations, in our case as
sums of three ${}_3F_2$ series, human insight and experience is needed --- not
to mention manipulatorial skills and computational perseverance. Consequently,
the question, whether it is possible to derive the recurrences \emph{directly
  from the integrals}, is a quite natural one.

We want to stress the point that the algorithmic theory is sufficiently
developed to carry out this task \emph{in principle}; namely, by applying
holonomic closure properties as introduced below. But for the integrals in
question the computational complexity of this approach turns out to be
prohibitively expensive. Nevertheless, there is an algorithmic workaround
which we describe in Section~\ref{sec.comput}. This workaround might be useful
also in other problems, and this is the reason why we wrote this short note.

The software we use is the package
\texttt{HolonomicFunctions}~\cite{Koutschan10b}, developed by the first-named
author in the computer algebra system Mathematica in the frame of his PhD
thesis~\cite{Koutschan09}. We start our investigations by loading this package
into the Mathematica system:
\begin{mma}
\In << |HolonomicFunctions.m| \\
\Print HolonomicFunctions package by Christoph Koutschan, RISC-Linz, Version 1.6 (12.04.2012) \\
\end{mma}

\section{The Holonomic Systems Approach}
\label{sec.holonom}

In order to state, in an algebraic language, the concepts that are introduced
in this section, and for writing mixed difference-differential equations in a
concise way, the following operator notation is employed: let $\D{x}$ denote
the partial derivative operator with respect to~$x$ ($x$ is then called a
\emph{continuous variable}) and $\S{n}$ the forward shift operator with respect to $n$
($n$ is then called a \emph{discrete variable}); they act on a function~$f$ by
\[
  \D{x}f = \frac{\d f}{\d x}
  \quad\text{and}\quad
  \S{n}f = f\big|_{n\to n+1}.
\]
They allow us to write linear homogeneous difference-differential equations
in terms of operators, e.g.,
\[
  \frac{\d}{\d x}f(k,n+1,x,y) + n\frac{\d}{\d y}f(k,n,x,y) + xf(k+1,n,x,y) - f(k,n,x,y) = 0
\]
turns into
\[
  \big(\D{x}\S{n} + n\D{y} + x\S{k} - 1\big) f(k,n,x,y) = 0;
\]
in other words, such equations are represented by polynomials in the operator
symbols $\D{x}$, $\S{n}$, etc., with coefficients in some field~$\F$ which we
assume to be of characteristic~$0$.  Typically, $\F$ is a rational function
field in the variables $x$, $n$, etc.  Note that in general the polynomial
ring $\F\langle\D{x},\S{n},\dots\rangle$ is not commutative (this fact is
indicated by the angle brackets) in the following sense: its coefficients from
$\F$ do not commute with the polynomial variables $\D{x}$, $\S{n}$, etc. For
instance, multiplication with $a(x,n)\in \F$ is subject to the rules
\[
  \D{x}\cdot a(x,n) = a(x,n)\cdot\D{x} + \frac{\d}{\d x} a(x,n)
  \quad\text{and}\quad
  \S{n}\cdot a(x,n) = a(x,n+1)\cdot\S{n}.
\]
Such non-commutative rings of operators are called \emph{Ore algebras},
denoted by~$\O$; concise definitions and specifications of the properties of
such algebras, for instance, can be found in~\cite{Koutschan09}.

We define the \emph{annihilator} (with respect to some Ore algebra~$\O$) of a
function~$f$ by:
\[
  \Ann_{\O}(f) := \{P\in\O\mid Pf=0\}.
\]
It can easily be seen that $\Ann_{\O}(f)$ is a left ideal in~$\O$. Every left
ideal~$I\subseteq\Ann_{\O}(f)$ is called an \emph{annihilating ideal} for~$f$.

\begin{definition}
  Let $\O=\F\langle\dots\rangle$ be an Ore algebra.  A function~$f$ is called
  \emph{$\d$-finite} w.r.t.~$\O$ if $\O/\Ann_{\O}(f)$ is a finite-dimensional
  $\F$-vector space. The dimension of this vector space is called the
  \emph{rank} of~$f$ w.r.t.~$\O$.
\end{definition}

In the holonomic systems approach, the representing data structures of
functions are (generators of) annihilating ideals (plus initial values). When
working with (left) ideals, we use \emph{(left) Gr\"obner
  bases}~\cite{Buchberger65,KandrirodyWeispfenning90} which are an important
tool for executing certain operations (e.g., the ideal membership test) in an
algorithmic way.

Without proof we state the following theorem about \emph{closure properties}
of $\d$-finite functions; its proof can be found in
\cite[Chap. 2.3]{Koutschan09}.  We remark that all of them are algorithmically
executable, and the algorithms work with the above mentioned data structure.

\begin{theorem}\label{thm.clprop}
Let $\O$ be an Ore algebra and let~$f$ and~$g$ be $\d$-finite w.r.t.~$\O$
of rank $r$ and $s$, respectively. Then
\renewcommand{\labelenumi}{(\roman{enumi})}
\setlength{\leftmargini}{2.5em}
\begin{enumerate}
\itemsep 0.5em
\item $f+g$ is $\d$-finite of rank $\leq r+s$.
\item $f\cdot g$ is $\d$-finite of rank $\leq rs$.
\item $f^2$ is $\d$-finite of rank $\leq r(r+1)/2$.
\item $Pf$ is $\d$-finite of rank $\leq r$ for any $P\in\O$.
\item $f|_{x\to A(x,y,\dots)}$ is $\d$-finite of rank $\leq rd$ if $x,y,\dots$
  are continuous variables and if the algebraic function $A$ satisfies a
  polynomial equation of degree~$d$.
\item $f|_{n\to A(n,k,\dots)}$ is $\d$-finite of rank $\leq r$ if $A$ is an
  integer-linear expression in the discrete variables $n,k,\dots$.
\end{enumerate}
\end{theorem}
\noindent
Note that in most examples the bounds on the rank are sharp. In
Section~\ref{sec.comput}, we exploit the fact that the rank does not grow when
applying closure properties (iv) or (vi).

\begin{example}\label{ex.Laguerre}
Consider the family of Laguerre polynomials $L_n^{(a)}(x)$ as an example of a
$\d$-finite function w.r.t. $\O=\Q(n,a,x)\langle\S{n},\S{a},\D{x}\rangle$.
The left ideal $I=\Ann_{\O}(L_n^{(a)}(x))$ is generated by the following three
operators that can be easily obtained with the \texttt{HolonomicFunctions} package:
\begin{mma}
  % Annihilator[LaguerreL[n, a, x], {S[n], S[a], Der[x]}]
  \In |Annihilator|[|LaguerreL|[n,a,x],\{|S|[n],|S|[a],|Der|[x]\}]\\
  \Out \{S_{\!a}+D_{\!x}-1,(n+1) S_{\!n}-x D_{\!x}+(-a-n+x-1),x D_{\!x}^2+(a-x+1) D_{\!x}+n\}\\
\end{mma}
These operators represent well-known identities for Laguerre polynomials.
Moreover, they are a left Gr\"obner basis of~$I$ with respect to the
degree-lexico\-gra\-phic order. Thus from the leading monomials ($\S{a}$, $\S{n}$,
and $\D{x}^2$) one can easily read off that the dimension of the
$\Q(n,a,x)$-vector space $\O/I$ is two, in other words: $L_n^{(a)}(x)$ is
$\d$-finite w.r.t. $\O$ of rank~$2$.
\end{example}

If we want to consider integration and summation problems, then the function
in question needs to be \emph{holonomic}, a concept that is closely related to
$\d$-finiteness. The precise definition is a bit technical and therefore
skipped here; the interested reader can find it, e.g.,
in~\cite{Zeilberger90,Coutinho95,Koutschan09}.  All functions that appear in
this paper are both $\d$-finite and holonomic. The following theorem
establishes the closure of holonomic functions with respect to sums and
integrals; for its proof, we once again refer
to~\cite{Zeilberger90,Koutschan09}.
\begin{theorem}
  Let the function $f$ be holonomic w.r.t. $\D{x}$ (resp. $\S{n}$). Then also
  $\int_a^b f\rmd x$ (resp. $\sum_{n=a}^b f$) is holonomic.
\end{theorem}
If a function is $\d$-finite and holonomic then Chyzak's
algorithm~\cite{Chyzak00} can be used to compute an annihilating ideal for the
integral (resp. sum), see Section~\ref{sec.ct}. In the following we apply this
algorithm to the Coulomb integrals presented in Section~\ref{sec.intro}.

\section{The Coulomb Integrals}
\label{sec.comput}

We now turn to the relativistic Coulomb integrals from Section~\ref{sec.intro}.
According to~\eqref{eq.FGdef} the wave functions are of the form
\[
  F = \big( \alpha_1L_{n-1} + \alpha_2L_n \big)E
  \quad\text{and}\quad
  G = \big( \beta_1L_{n-1} + \beta_2L_n \big)E
\]
where $L_n=L^{(2\nu)}_n(2a\beta r)$. Thus the expressions $F^2 \pm G^2$ that appear
in the integrands of~$A_p$ and $B_p$, respectively, can be written as follows:
\begin{equation}\label{eq.F2G2}
  F^2 \pm G^2 = \big( (\alpha_1^2 \pm \beta_1^2) L_{n-1}^2 +
    2(\alpha_1\alpha_2 \pm \beta_1\beta_2) L_{n-1}L_n +
    (\alpha_2^2 \pm \beta_2^2) L_n^2 \big)E^2.
\end{equation}
Similarly, for the integrand of $C_p$ we get
\begin{equation}\label{eq.FG}
  F\cdot G = \big( \alpha_1\beta_1L_{n-1}^2 + \alpha_2\beta_2L_n^2 \big)E^2
\end{equation}
since $\alpha_1\beta_2+\alpha_2\beta_1=0$ by \eqref{eq.alphas} and
\eqref{eq.betas}.  In this section we show how to derive linear
recurrence equations in~$p$ for the Coulomb integrals.

\subsection{Standard Closure Properties}
In order to treat the integral~$A_p$ with the holonomic systems approach, one
first has to transform the input, i.e., the integrand, into the required data
structure for $\d$-finite functions: given generators for the annihilating
ideals of $L_n$, $E$, and all other functions appearing in the right-hand side
of~\eqref{eq.F2G2}, an annihilating ideal for $F^2+G^2$ can be computed by the
closure properties addition, multiplication, and squaring. Considering only
the operator~$\D{r}$, Theorem~\ref{thm.clprop} (i,ii,iii) states that in this
case the rank of the result is at most
$(1\cdot3+1\cdot2\cdot2+1\cdot3)\cdot1=10$, since $L_n$ is of rank~$2$ (see
Example~\ref{ex.Laguerre}) and $E$ is hyperexponential in~$r$, i.e., satisfies
a first-order differential equation in~$r$. Recall that the remaining
coefficients are free of~$r$ and therefore also of rank~$1$.  It turns out
that the bound in this case is sharp, so that applying the closure property
algorithms implemented in \texttt{HolonomicFunctions} to the
expression~\eqref{eq.F2G2} yield an annihilating ideal of rank~$10$ which is
generated by a very large Ore polynomial in~$\D{r}$.  The situation is
exactly the same for $B_p$. For the integrand of $C_p$, the bound for the rank
is~$6$ by a similar reasoning.  Given these annihilating ideals as input, it
seems hopeless that the integration step via creative telescoping, see
Section~\ref{sec.ct}, can be completed in reasonable time.

\subsection{Annihilating Ideals for the Integrands}
Fortunately, there is a workaround as announced in the Introduction.  Namely,
we can find annihilating ideals of smaller rank by using different closure
properties: application of an operator (iv) and discrete substitution (vi) in
Theorem~\ref{thm.clprop}. We first demonstrate this idea on the expression
$\alpha_1\beta_1L_{n-1}^2 + \alpha_2\beta_2L_n^2$ that appears in $C_p$,
see~\eqref{eq.FG}. Instead of applying the closure property addition, this
expression can also be perceived as the operator
$\alpha_1\beta_1+\alpha_2\beta_2\S{n}$ applied to $L_{n-1}^2$.  As a
consequence of entry (iv) of Theorem~\ref{thm.clprop} one obtains an
annihilating ideal of rank~$3$, compared to rank~$6$ when closure properties
are employed in standard fashion. We start the computation by determining an
annihilating ideal of $L_{n-1}^2$:
\begin{mma}
  % annL2 = Annihilator[LaguerreL[n - 1, 2*\[Nu], 2*a*\[Beta]*r]^2, {S[n], Der[r], S[p]}]
  \In |annL2|=|Annihilator|[|LaguerreL|[n-1,2\nu,2a\beta r]\pow 2,\{|S|[n],|Der|[r],|S|[p]\}]\\
  \Out \{S_{\!p}-1,-r^2 D_{\!r}^2+2 n^2 S_{\!n}+(6 a \beta r^2-2 n r-6 \nu r-r) D_{\!r}+
    (-8 a^2 \beta ^2 r^2+4 a \beta n r+16 a \beta \nu r+4 a \beta r-8 \nu ^2-2 n^2-8 \nu n),
    -n r S_{\!n} D_{\!r}+2 n^2 S_{\!n}+(-n r-2 \nu r) D_{\!r}+
    (4 a \beta n r+8 a \beta \nu r-8 \nu ^2-2 n^2-8 \nu n),
    (n^3+2 n^2+n) S_{\!n}^2+(-4 a^2 \beta ^2 n r^2+8 a \beta n^2 r+8 a \beta \nu n r+
    4 a \beta n r-4 n^3-8 \nu n^2-4 n^2-4 \nu ^2 n-4 \nu n-n) S_{\!n}+
    (-2 a \beta n r^2-4 a \beta \nu r^2+2 n^2 r+6 \nu n r+n r+4 \nu ^2 r+2 \nu r) D_{\!r}+
    (8 a^2 \beta ^2 n r^2+16 a^2 \beta ^2 \nu r^2-12 a \beta n^2 r-40 a \beta \nu n r-
    4 a \beta n r-32 a \beta \nu ^2 r-8 a \beta \nu r+16 \nu ^3+8 \nu ^2+3 n^3+
    16 \nu n^2+2 n^2+28 \nu ^2 n+8 \nu n)\}\\
\end{mma}
Next, we have to apply the operator $\alpha_1\beta_1+\alpha_2\beta_2\S{n}$ to
the $\d$-finite function $L_{n-1}^2$. In order to keep the intermediate
expressions small, we replace the coefficients of the operator by simpler
ones: $c_1+c_2\S{n}$. Additionally, we divide the integrand by
$c_1=\alpha_1\beta_1$; this does not change the recurrence since $\alpha_1$
and $\beta_1$ depend neither on $r$ nor on~$p$, but we can get rid of one
parameter. The operator we want to apply to $L_{n-1}^2$ then reads
$1+q_2\S{n}$ with $q_2=\alpha_2\beta_2/(\alpha_1\beta_1)$. Still, the results
we get are somewhat large, so we suppress (by ending the input line with a
semicolon) the output of the following computations:
\begin{mma}
  % annFG = DFiniteOreAction[annL2, 1 + q2*S[n]];
  \In |annFG|=|DFiniteOreAction|[|annL2|,1+|q2|*|S|[n]];\\
\end{mma}
\noindent
To complete the derivation of an annihilating ideal for the integrand of~$C_p$,
we have to include the prefactor~$E$ (squared) and the additional factor $r^{p+2}$,
according to \eqref{eq.C} and~\eqref{eq.FG}:
\begin{mma}
  % prefactor = a^2*\[Beta]^(3/2)*Sqrt[n!/\[Gamma]/Gamma[n + 2*\[Nu]]]*(2*a*\[Beta]*r)^(\[Nu] - 1)*Exp[(-a)*\[Beta]*r];
  \In |prefactor|= \linebreak
    a\pow 2 \, \beta\pow (3/2) \, |Sqrt|[n!/\gamma/|Gamma|[n+2\nu]] \, (2a\beta r)\pow (\nu-1)|Exp|[-a\beta r];\\
  % annIntC = DFiniteTimes[Annihilator[prefactor^2*r^(p + 2), {S[n], Der[r], S[p]}], annFG];
  \In |annIntC|=|DFiniteTimes|[\linebreak
    |Annihilator|[|prefactor|\pow 2 * r\pow (p+2),\> \{|S|[n],|Der|[r],|S|[p]\}],\> |annFG|];\\
  % UnderTheStaircase[annIntC]
  \In |UnderTheStaircase|[|annIntC|]\\
  \Out \{1,D_{\!r},S_{\!n}\}\\
\end{mma}
\noindent
The last output shows that the rank is~$3$ (the number of monomials under the
staircase of the Gr\"obner basis), as expected.

Next we turn to the Coulomb integral~$A_p$, where the main part of its
integrand is given by~\eqref{eq.F2G2}.  Analogously to before, the key idea is
to rewrite the expression slightly as to interpret it as an operator applied
to some function, namely to the product of two Laguerre polynomials. The only
hurdle is that the indices of the Laguerre polynomials need to be shifted
separately: in order to produce $L_nL_{n-1}$ from $L_{n-1}L_{n-1}$, for
example, a mechanism is needed that shifts only the~$n$ in the first Laguerre
polynomial. This problem can be overcome by introducing a slack variable,
say~$m$, which afterwards is set to~$n$. The latter step is a discrete
substitution as it is described in part (vi) of Theorem~\ref{thm.clprop}, and
which corresponds to the computation of the diagonal of a bivariate
sequence. Thus one obtains
\[
  \frac{F^2+G^2}{E^2} = \big( (\alpha_1^2 + \beta_1^2) +
    2(\alpha_1\alpha_2 + \beta_1\beta_2)\S{n} + (\alpha_2^2 + \beta_2^2)\S{m}\S{n} \big)
  \big( L_{m-1}L_{n-1} \big)\Big|_{m\to n}
\]
and from Theorem~\ref{thm.clprop}, items (iv) and (vi), it is clear that the
rank of the corresponding annihilating ideal is at most~$4$; our computations
show that, once again, the bound is sharp. Similar to $C_p$ above, the
following commands yield an annihilating ideal for the integrand of $A_p$.
The only difference is that at the beginning we introduce the slack
variable~$m$ (and the corresponding operator~$\S{m}$), which later is
substituted by~$n$.  Again, we introduce new variables for the coefficients of
the operator in order to reduce the number of parameters: we use
$1+q_1\S{n}+q_2\S{m}\S{n}$ with
\begin{align}
q_1 & = 2(\alpha_1\alpha_2 + \beta_1\beta_2)/(\alpha_1^2 + \beta_1^2), \label{eq.q1} \\
q_2 & = (\alpha_2^2 + \beta_2^2)/(\alpha_1^2 + \beta_1^2). \label{eq.q2}
\end{align}
Changing all plus signs to minus signs gives the substitutions for $B_p$, so
that the result of the following calculations can be used both for $A_p$
and~$B_p$.
\begin{mma}
  % annLL = Annihilator[LaguerreL[m - 1, 2*\[Nu], 2*a*\[Beta]*r]*LaguerreL[n - 1, 2*\[Nu], 2*a*\[Beta]*r], {S[m], S[n], Der[r], S[p]}]
  \In |annLL|=|Annihilator|[\linebreak
    |LaguerreL|[m-1,2\nu,2a\beta r]| \, LaguerreL|[n-1,2\nu,2a\beta r],\linebreak
    \{|S|[m],|S|[n],|Der|[r],|S|[p]\}]\\
  \Out \{S_{\!p}-1,m S_{\!m}+n S_{\!n}-r D_{\!r}+(4 a \beta r-m-4 \nu -n),
    2 n r S_{\!n} D_{\!r}-r^2 D_{\!r}^2+(4 \nu n-4 a \beta n r) S_{\!n}+
    (6 a \beta r^2-2 n r-6 \nu r-r) D_{\!r}+(-8 a^2 \beta ^2 r^2-2 a \beta m r+
    6 a \beta n r+16 a \beta \nu r+4 a \beta r-8 \nu ^2-4 \nu n),
    (n+1) S_{\!n}^2+(2 a \beta r-2 \nu -2 n-1) S_{\!n}+(2 \nu +n),
    r^2 D_{\!r}^3+(-6 a \beta r^2+6 \nu r+3 r) D_{\!r}^2+
    (4 a \beta m n-4 a \beta n^2) S_{\!n}+(8 a^2 \beta ^2 r^2+2 a \beta m r+
    6 a \beta n r-16 a \beta \nu r-16 a \beta r+8 \nu ^2+6 \nu +1) D_{\!r}+
    (-16 a^2 \beta ^2 n r+16 a^2 \beta ^2 r-16 a \beta \nu -4 a \beta +
    2 a \beta m-4 a \beta m n+4 a \beta n^2+16 a \beta \nu n+2 a \beta n)\}\\
  % ann1 = DFiniteOreAction[annLL, 1 + q1*S[n] + q2*S[m]*S[n]];
  \In |ann1|=|DFiniteOreAction|[|annLL|,1+|q1|*|S|[n]+|q2|*|S|[m]*|S|[n]];\\
  % annF2G2 = DFiniteSubstitute[ann1, {m -> n}, Algebra -> OreAlgebra[Der[r], S[p]]];
  \In |annF2G2|=|DFiniteSubstitute|[|ann1|,\{m\to n\},\linebreak
    |Algebra|\to |OreAlgebra|[|Der|[r],|S|[p]]];\\
  % annIntA = DFiniteTimes[Annihilator[prefactor^2*r^(p + 2), {Der[r], S[p]}], annF2G2];
  \In |annIntA|=|DFiniteTimes|[\linebreak \label{mma.annIntA}
    |Annihilator|[|prefactor|\pow 2 * r\pow (p+2),\> \{|Der|[r],|S|[p]\}],\> |annF2G2|];\\
  % UnderTheStaircase[annIntA]
  \In |UnderTheStaircase|[|annIntA|]\\
  \Out \{1,D_{\!r},D_{\!r}^2,D_{\!r}^3\}\\
\end{mma}

\subsection{Creative Telescoping}
\label{sec.ct}
In the previous section, annihilating ideals for the three integrands in
\eqref{eq.A}--\eqref{eq.C} were derived. Taking these as input, the task is
now to compute annihilating ideals for the integrals themselves.  This goal
can be achieved by the method of \emph{creative
  telescoping}~\cite{Zeilberger91}.  To explain the key idea we restrict
ourselves to the situation we are confronted with, i.e., a single integral
with respect to $r$ which contains a discrete parameter~$p$ (but everything
can be stated in more general terms): find an operator~$P$ in the annihilating
ideal~$I$ of the integrand~$f(p,r)$ of the form
\begin{equation}\label{eq.ct}
  P(p,r,\S{p},\D{r}) = T(p,\S{p}) + \D{r}\cdot C(p,r,\S{p},\D{r}).
\end{equation}
The part~$T$ is called the \emph{telescoper} and $C$ is called the
\emph{certificate}.  Since $P\in I$ it follows that $P f = 0$; integrating
this equation and applying the fundamental theorem of calculus yields
\[
  T\left(\int_0^\infty f(p,r)\rmd r\right) + \left[C\big(f(p,r)\big)\right]_{r=0}^{r=\infty} = 0.
\]
This is the desired recurrence equation for the integral which, if necessary,
can be brought into homogeneous form; this is easily done in practice.  In
terms of annihilating ideals, we observe that $\Q(p)\langle\S{p}\rangle$ is a
principal ideal domain, and therefore the annihilator~$A$ of the integral is
generated by a single element. Trivially, the singleton $\{T\}$ is already a
left Gr\"obner basis. From the theory of holonomic modules~\cite{Coutinho95}
it follows that an operator of the form~$P$ always exists, provided the
integrand~$f$ is holonomic. But in general, it need not be the case that there
exists a $P\in I$ whose telescoper coincides with the unique generator
of~$A$. In other words, this method indeed succeeds in computing a recurrence
for the integral, but it may not be able to deliver one of \emph{minimal
  order}.

A first algorithm to compute creative telescoping operators was given
in~\cite{Zeilberger90}; it is based on elimination techniques and not very
efficient in practice.  Around the same time the
algorithms~\cite{Zeilberger90a,AlmkvistZeilberger90} were published; they are
more efficient, but restrict the input to terminating hypergeometric series
and hyperexponential functions, respectively.  These two algorithms were later
generalized in~\cite{Chyzak00}, and in~\cite{Koutschan10c} a heuristic
approach was presented which, in practice, completes the task very
quickly. All above-mentioned algorithms are implemented in the package
\texttt{HolonomicFunctions}~\cite{Koutschan10b}.  Concretely, we apply the
method~\cite{Koutschan10c} to the Coulomb integrals
\eqref{eq.A}--\eqref{eq.C}.

Starting with the annihilating ideal~$I$ (computed in the command line
In[\ref{mma.annIntA}] above) of the integrand of~$A_p$, the following command
computes operators $T$ and $C$ such that $T+\D{r}C$ is in~$I$.  The results
are too large to be printed here, so only their sizes are displayed.
\begin{mma}
  % {{annA}, {certA}} = FindCreativeTelescoping[annIntA, Der[r]];
  \In \{\{|annA|\},\{|certA|\}\}=|FindCreativeTelescoping|[|annIntA|,|Der|[r]];\\
  % ByteCount /@ {annA, certA}
  \In |ByteCount|/@\{|annA|,|certA|\}\\
  \Out \{75504,6682440\}\\
\end{mma}
\noindent
The following command gives the support of the computed operator in the
annihilating ideal~$A$ of the integral~$A_p$, which shows that the recurrence
for the integral is of order~$2$.
\begin{mma}
  % Support[annA]
  \In |Support|[|annA|]\\
  \Out \{S_{\!p}^2,S_{\!p},1\}\\
\end{mma}
The final step consists in replacing the temporarily introduced parameters by
their actual values. Using \eqref{eq.q1}, \eqref{eq.q2}, \eqref{eq.alphas},
\eqref{eq.betas}, and \eqref{eq.rel1}--\eqref{eq.rel3} one obtains
\begin{align*}
q_1 & = \frac{2\varepsilon\nu-2\varepsilon^2\kappa}{a\mu-\varepsilon\nu+\kappa},\\
q_2 & = \frac{-a\mu-\varepsilon\nu+\kappa}{a\mu-\varepsilon\nu+\kappa}.
\end{align*}
Futher simplifications of the recurrence with relations
\eqref{eq.rel1}--\eqref{eq.rel3} lead to the result given
in~\cite{Suslov10,PauleSuslov13}:
\begin{align*}
 A_{p+1} & = \frac{\mu\,P(p)}{a^2\beta\big(4\mu^2(p+1)+p(2\varepsilon\kappa+p)(2\varepsilon\kappa+p+1)\big)(p+2)} \> A_p - {} \\
 & \quad\ \frac{(4\nu^2-p^2)\big(4\mu^2(p+2)+(p+1)(2\varepsilon\kappa+p+1)(2\varepsilon\kappa+p+2)\big)p}
    {(2a\beta)^2\big(4\mu^2(p+1)+p(2\varepsilon\kappa+p)(2\varepsilon\kappa+p+1)\big)(p+2)} \> A_{p-1}
\end{align*}
where
\begin{align*}
 P(p) & = 2\varepsilon p(p+2)(2\varepsilon\kappa+p)(2\varepsilon\kappa+p+1) + {} \\
  & \quad\ \varepsilon\big(4\big(\varepsilon^2\kappa^2-\nu^2\big)-p\big(4\varepsilon^2\kappa^2+p(p+1)\big)\big) + {} \\
  & \quad\ (2p+1)\big(4\varepsilon^2\kappa+2(p+2)(2\varepsilon\mu^2-\kappa)\big).
\end{align*}
We do not show the calculations for the integrals $B_p$ and $C_p$ since they
can be done in an analogous way.

\section{Conclusion}
We studied the relativistic Coulomb integrals from the viewpoint of the
holonomic systems approach; in particular, we showed how these integrals can
be treated by computer algebra methods.  It is our hope that our exposition is
sufficiently instructive and puts the reader into the position to apply these
methods to his/her own benefit.  Additionally, we discussed several algorithmic
workarounds to reduce the complexity of the computations, as it turned out
that the integrals under consideration resist a naive application of our
software. We want to stress that the overall computing time for the integrals
\eqref{eq.A}--\eqref{eq.C} does not exceed a few minutes on a standard laptop.

Our next challenge is to study the off-diagonal matrix elements that are
important in applications
\cite{Puchkov10,Puchkov:Lab9,Puchkov:Lab10,Shab91,ShabHydVir} (see also the
references therein).  For the radial functions $F_{n, \kappa}(r)$ and $G_{n,
  \kappa}(r)$ given by (\ref{eq.FGdef}) in terms of the Laguerre polynomials,
one needs to investigate the following four integrals:
\begin{eqnarray*}
 && \int_0^\infty r^{p+2} \big(F_{n_1, \kappa_1} F_{n_2, \kappa_2} \pm G_{n_1, \kappa_1} G_{n_2, \kappa_2} \big) \rmd r, \\
 && \int_0^\infty r^{p+2} \big(F_{n_1, \kappa_1} G_{n_2, \kappa_2} \pm G_{n_1, \kappa_1} F_{n_2, \kappa_2} \big) \rmd r
\end{eqnarray*}
as off-diagonal extensions of (\ref{eq.A})--(\ref{eq.C}). The first-order
($4\times 4$ matrix) recurrence relations in $p$ among these integrals are
derived in~\cite{Puchkov10} from a virial theorem.

A straightforward consideration requires multiple evaluations of the following
integrals:
\[
  \int_{0}^{\infty} L_{n_{1}}^{(\lambda_{1})}(k_{1}x)  L_{n_{2}}^{(\lambda_{2})}(k_{2}x) \, x^{\mu+s}e^{-x}\rmd x,
\]
while it is almost impossible to simplify the lengthy end results. 
Computer algebra methods and, in particular, the holonomic systems
approach suggest another path which will be pursued elsewhere.

\subsection*{Acknowledgment}
The third-named author thanks the Research Institute for Symbolic Computation
(RISC) for the hospitality during his visit.

\bibliographystyle{plain}
%\bibliography{coulomb}

\end{document}